\documentclass[11pt]{article}

\usepackage{amssymb,amsmath,amsfonts,amsthm,lscape}

 \usepackage{cite}
 \usepackage{hyperref}
\hypersetup{
   colorlinks   =  true,
    linkcolor    = blue,
    citecolor    = red,
     urlcolor	=magenta,     
}

\newcommand\be{\begin{equation}}
\newcommand\ee{\end{equation}}
\newcommand\bea{\begin{eqnarray}}
\newcommand\eea{\end{eqnarray}}

 \def\dd{{\rm d}}
\def\eee{{\rm e}}
 \def\gam{\Theta}
\def\Gam{\Gamma}

\def\>#1{{\bf #1}}

\begin{document}

\thispagestyle{empty}

\

 \vskip1cm

\noindent {\Large{\bf {Generalized time-dependent SIS Hamiltonian models:\\[4pt] Exact solutions and quantum deformations
}}}

\medskip
\medskip
\medskip
\medskip

 \begin{center}

 {\sc Eduardo Fern\'andez-Saiz\footnote{
 Based on the contribution presented at ``The XII International Symposium on Quantum Theory and Symmetries" (QTS12),
July 24--28, 2023, 
 Czech Technical University in Prague, Czech Republic.\\ To appear in Journal of Physics: Conference Series.
}, Rutwig Campoamor-Stursberg$^{2,3}$\\[2pt] and Francisco  J.~Herranz$^4$}

 \end{center}

\medskip

\noindent
 {$^1$ Department of Quantitative Methods, CUNEF Universidad, 
C. de Leonardo Prieto Castro 2, E-28040  Madrid, Spain} \\[2pt]
\noindent
{$^2$ Instituto de Matem\'atica Interdisciplinar, Universidad Complutense de Madrid, E-28040 Madrid,  Spain
} \\[2pt]
{$^3$ Departamento de Geometr\'{\i}a y Topolog\'{\i}a,  Facultad de Ciencias 
Matem\'aticas, Universidad Complutense de Madrid, Plaza de Ciencias 3, E-28040 Madrid, Spain
} \\[2pt]
{$^4$ Departamento de F\'isica, Universidad de Burgos, 
E-09001 Burgos, Spain}
 
\medskip

 {E-mails:\quad {\tt  e.fernandezsaiz@cunef.edu, rutwig@ucm.es,  fjherranz@ubu.es}}

\begin{abstract}  
\noindent
The theory of  Lie--Hamilton  systems is used to construct generalized time-dependent SIS epidemic Hamiltonians with a variable infection rate from the `book' Lie algebra. Although these are characterized by  a set of non-autonomous nonlinear and coupled differential equations,  their corresponding exact solution is explicitly found. Moreover, the quantum deformation of the book algebra is also considered, from which the corresponding deformed 
SIS Hamiltonians are obtained and interpreted as perturbations in terms of the quantum deformation parameter of previously known SIS systems. The exact solutions for   these deformed systems are also obtained.

\end{abstract}


\section{Introduction}

In this contribution, we make use of the theory of Lie--Hamilton (LH) systems~\cite{CLS13, BCHLS13Ham, LH2015, BHLS, LuSa} in order to construct 
generalized time-dependent SIS epidemic models with a variable infection rate. We recall that  LH systems form  a particular class within Lie systems~\cite{LuSa, PW, CGM00, CGM07, CGL2010}, due to the existence of Hamiltonian vector fields with respect to a Poisson structure.

In particular,  we shall consider in the following the two-dimensional  book Lie algebra   $\mathfrak b_2={\rm span}\{ v_A, v_B\}$ with Lie bracket $[v_A,v_B] = - v_B$, corresponding  to the class I$_{14A}^{r=1}\simeq \mathbb{R} \ltimes \mathbb{R}$  in the classification of  LH systems on $\mathbb R^2$~\cite{LH2015,BHLS}.
It is thus ensured that any Lie system with two vector fields satisfying the above Lie bracket  belongs to this class in the classification and that the Lie system is  locally diffeomorphic  to  $\mathfrak b_2$. In this sense, we stress that the non-trivial issue   is to  obtain explicit local diffeomorphisms between LH systems belonging to the same class. It is worth mentioning that $\mathfrak{b}_2$-LH 
systems have already been studied in various mathematical, physical  and biological contexts such as 
\begin{itemize}

\item {\em Generalised Buchdahl equations}, i.e., second-order differential equations  arising  in the study of relativistic fluids~\cite{Buchdahl,Duarte,Chandrasekar},  
were shown in~\cite{LH2015,BHLS} to be  $\mathfrak{b}_2$-LH 
systems.   These  systems  have also been  studied from a Lagrangian approach in~\cite{Nikiciuk}.

\item Some   particular two-dimensional {\em  Lotka--Volterra systems   with $t$-dependent coefficients}~\cite{Tsvetkov,Jin} have been identified with $\mathfrak{b}_2$-LH 
systems in~\cite{LH2015,BHLS}.

\item {\em Complex Bernoulli differential equations with $t$-dependent real coefficients}~\cite{Muriel}; these  are particular cases of the non-autonomous complex Bernoulli differential equations with  complex coefficients~\cite{Zoladek, Marino}. In an LH framework, these have been analyzed in~\cite{Ballesteros6,BHLS,Bernoulli}.

  \item Quite recently, $\mathfrak{b}_2$-LH 
systems have also   been  shown to underlie the description of {\em generalized stochastic SIS epidemic models with   variable  infection rates}~\cite{Covid,CFH}, which is the case that we consider in the following.

\end{itemize}

In the next section, we construct the LH systems  coming from the book algebra in a `canonical' form. Furthermore, we obtain their    exact solution that depends on two arbitrary $t$-dependent coefficients. In section~\ref{s3}, these results are applied to $t$-dependent SIS models by deducing an appropriate explicit diffeomorphism (change of variables) with the canonical expressions. In particular,   we review the main results presented in~\cite{CFH}, thus   generalizing the SIS Hamiltonians previously studied in~\cite{Covid} and \cite{NakamuraMartinez}.

In section~\ref{s4} , we apply  the formalism of Poisson--Hopf   deformations of LH systems proposed in~\cite{Ballesteros6,BCFHL,BCFHLb} to the book algebra. We consider the quantum deformation of $\mathfrak{b}_2$ and obtain the corresponding deformed LH systems, deducing their exact solution. In section~\ref{s41}, as a new result that complements those of~\cite{CFH}, we obtain deformed SIS   differential equations together with their general solution.


\section{Lie--Hamilton systems from the book   algebra}
\label{s2}

Let us consider the so-called two-dimensional book Lie algebra $\mathfrak b_2={\rm span}\{ v_A, v_B\}$,   obeying the Lie bracket
$[v_A,v_B] = - v_B$. We can interpret $v_A$ as a  dilation  generator, while  $v_B$ corresponds to a translation one. As happens with any Lie algebra, $\mathfrak b_2$ can be endowed with a trivial Hopf algebra structure~\cite{CP,Abe} with (primitive) coproduct $\Delta : \mathfrak b_2 \to \mathfrak b_2 \otimes \mathfrak b_2 $ given by
\be
\Delta (v_i)=v_i \otimes 1 + 1\otimes v_i  ,\qquad i=A,B.
\label{b2}
\ee
For any  algebra $\mathcal A$, this map   must be an algebra homomorphism and satisfy the coassociativity condition in $\mathcal A\otimes \mathcal A \otimes \mathcal A$:
\be
({\rm Id}\otimes\Delta)\Delta(a)=(\Delta\otimes {\rm Id})\Delta(a),\qquad \forall a\in \mathcal A ,
\label{b2a}
\ee
such that the pair $(\mathcal A,\Delta)$ defines a coalgebra.
These properties are trivially fulfilled for any Lie algebra with non-deformed coproduct (\ref{b2}).

We  introduce a symplectic representation $D$ of $\mathfrak b_2$ in terms of canonical variables $(x,y)$ and the standard symplectic form
$\omega = \dd x\wedge  \dd y$
 defined by~\cite{Ballesteros6}
\be
h_A:= D(v_A)=  xy,\qquad h_B:= D(v_B)=  - x, 
\label{b4}
\ee
where the Hamiltonian functions  fulfill the following Poisson bracket (with respect  to $\omega$):
\be
\{ h_A,h_B\}_\omega= - h_B .
\label{b5}
\ee
Using (\ref{b4}), we can easily deduce a differential representation of $\mathfrak b_2$ in terms of Cartesian coordinates $(x,y)\in \mathbb R^2$ through the contraction or inner product operation,  
$\iota_{{\bf X}_i}\omega={\rm d}h_i$, that provides the following vector fields
 \be
     \>X_A=x\, \frac{\partial}{\partial x}- y\, \frac{\partial}{\partial y}, \qquad    \>X_B=\frac{\partial}{\partial y},\qquad  [\>X_A,\>X_B]=\>X_B .
\label{b7}    
\ee
By construction, these verify the invariance condition for  the Lie derivative:
${\cal L}_{\mathbf{X}_i}\omega =0$.

From the covariant or contravariant approach, we obtain either a $t$-dependent Hamiltonian or a $t$-dependent vector field depending on two real arbitrary parameters $b_A(t)$ and $b_B(t)$:
\begin{equation} 
\begin{split}
h_t&=b_A(t)  h_A +b_B(t)  h_B   =  b_A(t)\,  xy - b_B(t) \,x   , \\
  \>X_t&=b_A(t)   \>X_A +b_B(t)   \>X_B   =  b_A(t) \left( x\, \frac{\partial}{\partial x}- y\, \frac{\partial}{\partial y} \right)    + b_B(t) \, \frac{\partial}{\partial y}  ,
\end{split}
 \label{b8}
\end{equation}
that both  lead to the same system of non-autonomous  ODEs on $\mathbb R^2$:
   \begin{equation} 
\frac{\dd x}{\dd t}=b_A(t)\,  x ,
\qquad
\frac{\dd y}{\dd t}=-b_A(t)\, y+ b_B(t)  .
 \label{b9}
\end{equation}

It is straightforward to verify that, due to the relation (\ref{b7}), the ODEs (\ref{b9}) possess the structure of  a Lie system~\cite{PW,CGM00,CGM07,CGL2010,LuSa} with associated $t$-dependent vector field $ \>X_t$  (\ref{b8}) and
Vessiot--Guldberg Lie algebra isomorphic to $\mathfrak b_2$. By  the Lie--Scheffers Theorem,  this guarantees that the system always admits a fundamental system of solutions, that is, a superposition rule.

\smallskip
As   (\ref{b9}) is separable in the coordinates $(x,y)$, the latter property is not required, and the system can be solved explicitly by quadratures, with the exact solution given by 
  \be
\begin{split}
x(t)=c_1 \,\eee^{\gam(t)} ,\qquad
y(t)=\left( c_2 +  \int_a^t  \eee^{\gam(u )} b_B(u) \dd u\right)  \eee^{-\gam(t)} ,\qquad \gam(t):= \int_a^t b_A(s)\dd s ,
\end{split}
\label{b10}
\ee
where $c_1$ and $c_2$ are the two  constants of integration, determined by the  initial conditions, while $a$ is a real number that ensures the existence of the integrals over the interval $[a,t]$. Now, for any (local) diffeomorphism, i.e., a change of variables, such that a $\mathfrak{b}_2$-LH  system is expressed in variables different from $(x, y)$, the general solution can be derived directly using formula (\ref{b10}). 
 This strategy was used  in~\cite{CFH} in order to obtain   exact solutions for  generalized stochastic SIS epidemic models, which, in the proper basis, correspond to a set of non-autonomous nonlinear and coupled differential equations.  Additionally,   the very same procedure has also been applied to complex Bernoulli differential equations with $t$-dependent real coefficients  deducing their exact solution in~\cite{Bernoulli}.


 \section{Applications  to generalized time-dependent SIS Hamiltonian models}
\label{s3}

In this section we briefly review the main results concerning the application of the `abstract' $\mathfrak{b}_2$-LH algebra described in section~\ref{s2} to Hamiltonian systems that generalize  SIS epidemic models with a time-dependent infection rate (see~\cite{CFH} for details).


\subsection{A time-independent SIS Hamiltonian model}
\label{s31}

The starting point is the SIS model introduced by Nakamura and Martinez in~\cite{NakamuraMartinez}. For more details on the use of differential equations to study the population dynamics of infectious diseases we refer to~\cite{Bai,Hethcote,Walt,KK,Ab52,Miller}. In these models,  all individuals are still susceptible to the infection after recovery, which means that they do not acquire immunity. Therefore, the model can be described only using two compartmental variables. The first one, `I', corresponds to the number of infected individuals, meanwhile the second compartment `S' describes the number of individuals susceptible to the infection at a given time.

Under certain assumptions~\cite{CFH,NakamuraMartinez}, the differential equation that determines  the density $\rho=\rho(t)$ of infected individuals  can be written as
\begin{equation}
\frac{\dd\rho}{\dd t}= \rho(\rho_0-\rho),
\label{sismodel0}
\end{equation}
where $\rho_0$ is a constant that corresponds to the equilibrium density. Then, following~\cite{NakamuraMartinez}, we introduce    fluctuations in the SIS model  in such a manner  that the spreading of the disease is interpreted as a Markov chain in discrete time, with at most one single recovery or transmission occurring in each infinitesimal interval. By   introducing   the   mean density of infected individuals $\langle\rho\rangle$ and the variance $\sigma^2=\langle\rho^2\rangle-\langle\rho\rangle^2$~\cite{kiss},  assuming that $\sigma$ becomes negligible when compared to $\langle\rho\rangle$ and ignoring higher statistical moments,   for a sufficiently large number of individuals  we are led to the system 
\begin{equation}\label{sislabellog}
\begin{split}
\frac{\dd \ln{\langle\rho\rangle}}{\dd t}=\rho_0-\langle\rho\rangle-\frac{\sigma^2}{\langle\rho\rangle},\qquad
\frac{1}{2}\frac{\dd \ln{\sigma^2}}{\dd t}=\rho_0-2\langle\rho\rangle ,
\end{split}
\end{equation}
which  corresponds to a stochastic expansion, according to \cite{vilar}. In this situation, we assume that the density $\rho=\langle \rho\rangle +\eta$ is properly described by the instantaneous average, as well as some noise function $\eta$. For consistency, it is  further assumed that $\langle \eta\rangle = 0$ and $\langle \eta^2\rangle = \sigma^2$. The   system \eqref{sislabellog} allows a Hamiltonian formulation, with the phase space variables given by the mean  density of infected individuals $\langle\rho\rangle$ and the variance $\sigma^2$. Defining 
  \be
  q:=\langle\rho\rangle, \qquad p:=\sigma^{-1}
  \label{z1}
  \ee
   as  dynamical variables~\cite{NakamuraMartinez}, the system \eqref{sislabellog} now adopts the form 
\begin{equation}\label{sismodel3}
\begin{split}
\frac{\dd q}{\dd t}=\rho_0q-q^2-\frac{1}{p^2},
\qquad
\frac{\dd p}{\dd t}=-\rho_0p+2qp.
\end{split}
\end{equation}
These are the canonical equations associated to the Hamiltonian given by 
\begin{equation}\label{hamiltoniansis}
H=qp\left(\rho_0-q\right)+\frac{1}{p}.
\end{equation}
Both the systems (\ref{sismodel3}) and (\ref{sislabellog}) can be explicitly integrated; the exact solution and a  
 detailed analysis of all of the above results can be found in~\cite{NakamuraMartinez}.


\subsection{Time-dependent SIS Hamiltonian models}
\label{s32}

A generalization of the Hamiltonian (\ref{hamiltoniansis}) was proposed in~\cite{Covid}, by considering a $t$-dependent      infection rate through a smooth function $\rho_0(t)$. The latter amounts to introduce a $t$-dependent basic reproduction number $R_0(t)$, a fact that is actually observed in more accurate epidemic models~\cite{Cui,Gao,Driessche}. The remarkable point  is that the resulting $t$-dependent Hamiltonian inherits the structure of an LH system~\cite{CLS13,BCHLS13Ham, LH2015, BHLS, LuSa}. Furthermore,   it is possible to extend such 
$t$-dependent SIS Hamiltonian by considering a second arbitrary $t$-dependent parameter $b(t)$ in the form~\cite{CFH}
\begin{equation}
h_t=\rho_0(t)h_A+b(t) h_B  ,\qquad h_A=qp,\qquad h_B= \frac{1-q^2p^2}{p},
 \label{se}
\end{equation}
leading to the following system of differential equations:
\begin{equation} 
\begin{split}
\frac{\dd q}{\dd t}=\rho_0(t) q - b(t) \left(  q^2+\frac{1}{p^2} \right)  ,
\qquad
\frac{\dd p}{\dd t}=-\rho_0(t) p+2b(t)  qp.
\end{split}
 \label{sf}
\end{equation}
The associated  $t$-dependent vector field is given by 
\be
{\bf X}_t=\rho_0(t){\bf X}_A+b(t){\bf X}_B, \quad\  {\bf X}_A=q\,\frac{\partial}{\partial q}-p\,\frac{\partial}{\partial p},\quad\ {\bf X}_B=-\left(q^2+\frac{1}{p^2}\right)\frac{\partial}{\partial q}+2 q p\,\frac{\partial}{\partial p},
 \label{sg}
\ee
 fulfilling the commutation rule (\ref{b7}). We conclude that, for any choice of the parameters $\rho_0(t)$ and $ b(t)$, the  differential equations (\ref{sf})  determine  an LH system  with  associated    $\mathfrak{b}_2$-LH algebra, including the SIS epidemic models studied in~\cite{Covid} for  $b(t)\equiv 1$, as well as the model (\ref{sismodel3}) introduced in~\cite{NakamuraMartinez} for   $\rho_0(t)\equiv \rho_0$ and $b(t)\equiv 1$.

The classification of LH systems on the plane~\cite{LH2015,BHLS} implies that there must exist a 
  canonical transformation between the variables $(q,p)$ in (\ref{sf}) and  $(x,y)$ in (\ref{b9}).  This reads as (see~\cite{CFH})
  \be
\begin{split}
x=\frac {q^2 p^2-1} p ,\qquad y= \frac{qp^2}{q^2p^2-1},\qquad
 q=\frac{x^2 y}{x^2 y^2 -1} , \qquad  p = \frac{x^2  y^2 -1}x,
\end{split}
\label{sk}
\ee
while keeping  the canonical symplectic form  $\omega = \dd x \wedge\dd y= \dd q \wedge\dd p $. Observe 
that   the change of variables considered in~\cite{Covid}  is not canonical, so that  the variables $q$ and $p$ used there  do not correspond to    the mean  density of infected individuals $\langle\rho\rangle$ and the variance $\sigma^2$ (\ref{z1}).

Taking into account the results presented in section~\ref{s2}, we directly obtain the exact solution for the   book SIS Hamiltonian model (\ref{sf}).  Explicitly, if we start with  the general solution for any book LH system (\ref{b10}),  apply the transformation (\ref{sk}),  and identify the `abstract' $t$-dependent coefficients as   $b_A(t)\equiv \rho_0(t)$ and   $b_B(t)\equiv b(t)$, we find that 
    \be
\begin{split}
q(t)&=\frac{\left( c_2 +  \int_a^t  \eee^{\gam(u )} b(u) \dd u\right)\eee^{\gam(t)} }{\left( c_2 +  \int_a^t  \eee^{\gam(u )} b(u) \dd u\right)^2-c_1^{-2} } \, ,\qquad \gam(t):= \int_a^t\rho_0(s)\dd s , \qquad  \\ 
p(t)&=\left(  c_1\biggl( c_2 +  \int_a^t  \eee^{\gam(u )} b(u) \dd u\biggr)^2- c_1^{-1}  \right)  \eee^{-\gam(t)}  .
\end{split}
\label{sn}
\ee
Hence,  the exact solution of the $t$-dependent SIS system  considered in~\cite{Covid} arises by simply setting $b(t)\equiv 1$ and keeping a variable $\rho_0(t)$  which, in fact, was not presented in that work.
Moreover, if we fix  $b(t)\equiv 1$, consider a constant  $\rho_0(t)\equiv \rho_0$, and  choose $a=0$ in the integral $\gam(t)$ (i.e.~$\gam(t)= \rho_0 t$), we recover   the solution of the initial $t$-independent SIS model (\ref{sismodel3}), namely
     \be
q(t)=\frac{\rho_0 \left(   \eee^{\rho_0 t}+ c_2\rho_0-1\right)\eee^{\rho_0 t} }{ \left(   \eee^{\rho_0 t}+ c_2\rho_0-1\right)^2-\rho_0^2c_1^{-2} },\qquad 
p(t)=\left(     \frac{ c_1\left(   \eee^{\rho_0 t}+ c_2\rho_0-1\right)^2  } {\rho_0^2}  - \frac 1{c_1}   \right)  \eee^{-\rho_0 t}   .
\label{so}
\ee
 

\section{Deformed Lie--Hamilton systems from the quantum   book\\   algebra}
\label{s4}

In this section, we apply the formalism of Poisson--Hopf   deformations of LH systems proposed in~\cite{Ballesteros6,BCFHL,BCFHLb} to the book algebra. This will allow us to obtain all deformed LH systems coming from a quantum deformation of $\mathfrak{b}_2$. We will later particularize these results  to  obtain novel deformed SIS   differential equations, that by the preceding discussion admit an exact solution. 

We  consider the coboundary quantum deformation of the book Lie algebra $\mathfrak{b}_2$    coming from the classical $r$-matrix
$r=z  v_A\wedge v_B$,  which is a solution of the classical Yang--Baxter equation, and  where $ z$ is the quantum deformation parameter such that $q={\rm e}^z$. For  quantum deformations of the three-dimensional book algebra $\mathfrak{b}_3$, we refer to~\cite{Ballesteros2012}. The quantum book  algebra $U_{   z}(\mathfrak{b}_2)\equiv \mathfrak{b}_{z,2 }$, is  defined by the following deformed coproduct, fulfilling the coassociativity  property (\ref{b2a}), and  compatible deformed commutation relation:
 \be
\begin{split}
\Delta_z(v_A)&=v_A\otimes \eee^{-z v_B}+1\otimes v_A , \quad\ 
\Delta_z(v_B)=v_B\otimes 1 + 1 \otimes v_B , \quad\
[v_A,v_B]_z= \frac{\eee^{-z v_B}-1}{z} .
\end{split}
\label{d5}
\ee

A deformed     symplectic representation $D_z$ of $\mathfrak b_{z,2}$  (\ref{d5}), in terms of  the canonical variables $(x,y)$ of section~\ref{s2} and keeping the canonical symplectic form $\omega$, is given by 
 \be
h_{z,A}:= D_z(v_A)=  \left( \frac{ \eee^{z x} -1}{z}\right) y,\qquad h_{z,B}:= D_z(v_B)=  - x ,
\label{d7}
\ee
with the following deformed   Poisson bracket with respect  to $\omega$:
\be
\{h_{z,A},h_{z,B}\}_\omega=\frac{\eee^{-z h_{z,B}}-1}{z}  .
\label{d8}
\ee
The corresponding vector fields turn out to be
\be
  \>X_{z,A}=\left( \frac{ \eee^{z x}-1}{z} \right)\frac{\partial}{\partial x}- \eee^{z x} y \, \frac{\partial}{\partial y}   ,
\qquad  \>X_{z,B}= \frac{\partial}{\partial y}   ,
\label{d9}
\ee
 spanning a smooth distribution in the sense of  Stefan--Sussmann~\cite{Vaisman,Pa57,WA} through the commutator
 \be
 [{\bf X}_{z,A},{\bf X}_{z,B}]=  \eee^{z x} \, {\bf X}_{z,B} .
 \label{d10}
 \ee
 
Hence, we arrive at   a deformed $t$-dependent Hamiltonian and   vector field depending on two real arbitrary parameters $b_A(t)$ and $b_B(t)$ as follows (compare with (\ref{b8})):
\begin{equation} 
\begin{split}
h_{z,t}&=b_A(t)  h_{z,A} +b_B(t)  h_{z,B}   =  b_A(t)   \left( \frac{ \eee^{z x} -1}{z}\right) y - b_B(t) \,x   , \\
  \>X_{z,t}&=b_A(t)   \>X_{z,A} +b_B(t)   \>X_{z,B}   =  b_A(t) \left( \left( \frac{ \eee^{z x}-1}{z} \right)\frac{\partial}{\partial x}- \eee^{z x} y \, \frac{\partial}{\partial y} \right)    + b_B(t) \, \frac{\partial}{\partial y}  .
\end{split}
 \label{d11}
\end{equation}
Either one gives rise to the following system of non-autonomous nonlinear and coupled  ODEs on $\mathbb R^2$  
\begin{equation} 
\begin{split}
\frac{\dd x}{\dd t}=  b_A(t) \left( \frac{ \eee^{z x}-1}{z} \right)   , \qquad
\frac{\dd y}{\dd t}= - b_A(t) \eee^{z x}   y+b_B(t)   ,
\end{split}
 \label{d12}
\end{equation}
generalizing (\ref{b9}). Therefore,  this result can be seen as the general `canonical'  system of differential equations for  the set of deformed LH systems based in the quantum  book algebra $\mathfrak b_{z,2}$~(\ref{d5}).

Observe that the expressions (\ref{d7})--(\ref{d12}) reduce to (\ref{b4})--(\ref{b9})  under the (classical) non-deformed  limit $z\to 0$.
The presence of the quantum deformation parameter $z$  can be regarded as the introduction of a perturbation in  the initial LH system (\ref{b9}), in such a manner that  a nonlinear interaction or coupling between the  variables $(x,y)$ in the deformed LH system (\ref{d12})   arises  through the term $   \eee^{z x} y$. This fact can be clearly appreciated 
by taking a power series expansion in $z$ of  (\ref{d12}) and truncating at the first-order, namely
\begin{equation} 
\begin{split}
\frac{\dd x}{\dd t}= b_A(t)  \bigl(  x  +\tfrac 12 z    x^2 \bigr) +o[z^2]  , \qquad
\frac{\dd y}{\dd t}=  - b_A(t) \bigl(   y +z\,     x y  \bigr)+b_B(t) +o[z^2]   ,
\end{split}
 \label{d13}
\end{equation}
which holds for a small value of   $z$. In this approximation, we find that   $z$ introduces a quadratic term $x^2 $ into the first equation,  so becoming a standard Bernoulli  equation, while it introduces a  nonlinear interaction with the $xy$  term 
  in the second equation.

The exact solution  for (\ref{d12}) can be derived by solving the first differential equation, in which only  $x$ is involved, and then substituting this result into the second equation,  yielding (see~\cite{Bernoulli})
   \be
\begin{split}
x(t)&=-\frac{\ln \left(1- z c_1 \,\eee^{\gam(t)} \right)}z   ,\qquad \gam(t):= \int_a^t b_A(s)\dd s ,\\
y(t)&=  \exp\left( -\int_a^t \frac {b_A(u)}{1- z c_1\,\eee^{\gam(u)}  } \, \dd u\right)  \left( c_2 +  \int_a^t     \exp\left(  \int_a^u \frac {b_A(v)}{1- z c_1\,\eee^{\gam(v)}  } \, \dd v\right)  b_B(u) \dd u \right)  ,
\end{split}
\label{d14}
\ee
where, as in (\ref{b10}),  $c_1$ and $c_2$ are the two  constants of integration and $a$ is a real number that ensures the existence of the integrals over the interval $[a,t]$.  We stress that this  result  encompasses the solution of the quantum deformation of all the specific $\mathfrak b_{2}$-LH systems mentioned in the introduction, provided that a proper diffeomorphism/change of variables is known.


 \subsection{Applications  to  deformed time-dependent SIS Hamiltonian models}
\label{s41}

 As a new result, not covered in~\cite{CFH}, we apply the above quantum deformation to the $t$-dependent SIS Hamiltonian (\ref{se}), obtaining its  generalization in terms  of the quantum deformation parameter $z$.  We introduce the canonical change of variables (\ref{sk}), which preserves the interpretation (\ref{z1}), into the deformed Hamiltonian vector fields (\ref{d7}), finding that
 \be
   h_{z,A}=\frac {qp^2}{z\bigl(q^2 p^2 -1\bigr)}\left(   \exp\left(z\, \frac{q^2 p^2 -1}{p} \right)-1 \right),\qquad h_{z,B}= \frac{1-q^2p^2}{p} .
 \label{x1}
\ee
The deformed Hamiltonian (\ref{d11}) with $b_A(t)\equiv \rho_0(t)$ and   $b_B(t)\equiv b(t)$ leads to the   system of differential equations given by
\begin{equation} 
\begin{split}
\frac{\dd q}{\dd t}&=\rho_0(t) q\,\frac{2 p -\bigl(2p + z - z q^4p^4 \bigr) \exp\!\left(z\, \frac{q^2 p^2 -1}{p} \right)}{z\bigl(q^2 p^2 -1\bigr)^2} - b(t) \left(  q^2+\frac{1}{p^2} \right)  ,
\\
\frac{\dd p}{\dd t}&=-\rho_0(t) p^2 \,\frac{1+q^2 p^2 -\bigl(1+ q^2 p^2    + 2z q^2 p(1 -  q^2p^2 )\bigr) \exp\!\left(z\, \frac{q^2 p^2 -1}{p} \right)}{z\bigl(q^2 p^2 -1\bigr)^2} +2b(t)  qp.
\end{split}
 \label{x2}
\end{equation}
As a byproduct, this system covers the quantum deformation of the  SIS Hamiltonian    studied in~\cite{Covid} for  $b(t)\equiv 1$, together with the deformed counterpart of  (\ref{sismodel3}) given in~\cite{NakamuraMartinez} for   $\rho_0(t)\equiv \rho_0$ and $b(t)\equiv 1$. The associated deformed vector fields $\> X_{z,i}$ can  be easily deduced from (\ref{x2}).

 The exact solution of the system (\ref{x2}) is explicitly deduced from the change of variables (\ref{sk})
and the solution (\ref{d14}) for   deformed book systems in the coordinates $(x,y)$; the precise expressions are given by 
\begin{equation} 
\begin{split}
q(t)&=  \frac{ \eee^{-\Gam(t)} \left( c_2 +  \int_a^t     \eee^{\Gam(u)}  b(u) \dd u \right)  }{\eee^{-2\Gam(t)} \left( c_2 +  \int_a^t     \eee^{\Gam(u)}  b(u) \dd u \right)^2- z^2 \ln^{-2} \left(1- z c_1 \,\eee^{\gam(t)} \right)} ,
\\
p(t)&=\frac z{\ln \left(1- z c_1 \,\eee^{\gam(t)} \right)} - \frac{\ln \left(1- z c_1 \,\eee^{\gam(t)} \right)}z \, \eee^{-2\Gam(t)}\left( c_2 +  \int_a^t     \eee^{\Gam(u)}  b(u) \dd u \right)^2 ,\\
\gam(t)&:= \int_a^t \rho_0(s)\dd s , \qquad \Gam(t):= \int_a^t \frac {\rho_0(\tau)}{1- z c_1\,\eee^{\gam(\tau)}  } \, \dd \tau .
\end{split}
 \label{x3}
\end{equation}

Despite the cumbersome expressions for the novel deformed SIS system (\ref{x2}) and its exact solution  (\ref{x3}), one 
 can always interpret $z$ as a `small' perturbation parameter, as in~(\ref{d13}), which yields a  first-order $z$-perturbation of the $t$-dependent SIS system (\ref{sf}) in the form
\be
\begin{split}
\frac{\dd q}{\dd t}\simeq \rho_0(t) q - b(t) \biggl( \! q^2+\frac{1}{p^2}\! \biggr) + z\rho_0(t) q^3 p ,
\quad\
\frac{\dd p}{\dd t}\simeq2b(t)  qp-\rho_0(t) p+ \frac z2\, \rho_0(t) (1- 3 q^2 p^2).
\end{split}
 \label{x4}
\ee

To end with, it is worth stressing that the  same approach followed here for    $t$-dependent SIS models with a $\mathfrak{b}_2$-LH algebra symmetry, and its Poisson--Hopf deformation $\mathfrak{b}_{z,2}$, can be applied to other book LH systems such as  generalized Buchdahl equations~\cite{Buchdahl,Duarte,Chandrasekar,Nikiciuk} and Lotka--Volterra systems with $t$-dependent coefficients~\cite{Tsvetkov,Jin}, as well as certain higher dimensional algebras that contain $\mathfrak{b}_2$, such as the oscillator algebra $\mathfrak{h}_4$, for which exact solutions can also be derived, along the same lines. A detailed case by case analysis of exact solutions for these systems and their applications is currently in progress.



\section*{Acknowledgments}

R.C.S.~and F.J.H.~have been partially supported by Agencia Estatal de Investigaci\'on (Spain)  under grant  PID2019-106802GB-I00/AEI/10.13039/501100011033.  F.J.H.~acknowledges support  by the Regional Government of Castilla y Le\'on (Junta de Castilla y Le\'on, Spain) and by the Spanish Ministry of Science and Innovation MICIN and the European Union NextGenerationEU  (PRTR C17.I1). The authors also acknowledge the contribution of  RED2022-134301-T  funded by MCIN/AEI/10.13039/501100011033 (Spain).



\end{document}